# Real-time physics-informed reconstruction of transient fields using sensor guidance and higher-order time differentiation

Hong-Kyun Noh[1], Jeong-Hoon Park[1], Minseok Choi[2]*, and Jae Hyuk Lim[3]*

[1] *Department of Mechanical Engineering, Jeonbuk National University, 567 Baekje-daero, Deokjin-gu, Jeonju-si, Jeollabuk-do 54896, Republic of Korea*
[2] *Department of Mathematics, Pohang University of Science and Technology, 77 Cheongam-ro, Nam-gu, Pohang-si, Gyeongsangbuk-do 37673, Republic of Korea*
[3] *Department of Mechanical Engineering, Kyung Hee University, 1732 Deogyeong-daero, Giheung-gu, Yongin-si, Gyeonggi-do 17104, Republic of Korea*

## Abstract

This study proposes FTI-PBSM (Fixed-Time-Increment Physics-informed neural network-Based Surrogate Model), a novel physics-informed surrogate modeling framework designed for real-time reconstruction of transient responses in time-dependent Partial Differential Equations (PDEs) using only sparse, time-dependent sensor measurements.

Unlike conventional Physics-Informed Neural Network (PINN)-based models that rely on Automatic Differentiation (AD) over both spatial and temporal domains and require dedicated causal network architectures to impose temporal causality, the proposed approach entirely removes AD in the time direction. Instead, it leverages higher-order numerical differentiation methods, such as the Central Difference, Adams–Bashforth, and Backward Differentiation Formula, to explicitly impose temporal causality. This leads to a simplified model architecture with improved training stability, computational efficiency, and extrapolation capability.

Furthermore, FTI-PBSM is trained on sparse sensor measurements from multiple PDE cases generated by varying PDE coefficients, with the sensor data serving as model input. This enables the model to learn a parametric PDE family and generalize to unseen physical cases, accurately reconstructing full-field transient solutions in real time.

The proposed model is validated on four representative PDE problems—the convection equation, diffusion–reaction dynamics, Korteweg–de Vries (KdV) equation, and Allen–Cahn equation—and demonstrates superior prediction accuracy and generalization performance compared to a causal PBSM, which is used as the baseline model, in both interpolation and extrapolation tasks. It also shows strong robustness to sensor noise and variations in training data size, while significantly reducing training time.

* Corresponding authors: jaehyuklim@khu.ac.kr (J. H. Lim) and mchoi@postech.ac.kr (M. Choi)

# 1. Introduction

Many engineering and physical systems are governed by time-dependent Partial Differential Equations (PDEs), which describe the dynamic evolution of quantities such as displacement, temperature, concentration, or velocity. Solving these PDEs in real time is essential for applications in structural health monitoring, digital twin systems, and active control. However, conventional numerical methods such as the Finite Element Method (FEM) [1], Finite Difference Method (FDM) [2], and Finite Volume Method (FVM) [3], while accurate and robust, remain computationally prohibitive for real-time simulation, particularly in high-fidelity or nonlinear problems.

To address these limitations, Machine Learning (ML)-based surrogate models have emerged as attractive alternatives. Among them, Physics-Informed Neural Networks (PINNs) [4] have gained significant attention, as they embed the governing equations directly into the loss function. More recent studies have demonstrated that PINNs can be used to reconstruct full spatio-temporal fields from sparse sensor data [5-8], highlighting their potential for real-time inference with limited measurements. While these approaches have shown promise, particularly in dynamic systems governed by time-dependent PDEs, PINNs still face three well-documented limitations in such settings: (i) high cost from Automatic Differentiation (AD) in time, (ii) sensitivity to heuristic causal weightings or stage-wise solvers, and (iii) a tendency to converge to trivial or over-smoothed solutions when steep gradients or nonlinear waves are present.

To mitigate these issues, a variety of enhanced PINNs have been proposed. Representative examples include causal weightings that enforce temporal causality [9], Runge-Kutta PINNs (RK-PINNs) for stable higher-order time integration [10, 11], domain-decomposition schemes with hard constraints such as AT-PINN-HC [12], adaptive collocation that redistributes training points to large-residual regions [13, 14], and gradient-enhanced PINNs (gPINNs) that attach residual gradients to the loss for higher accuracy with fewer collocation points [15]. These refinements improve stability and convergence, yet they still require backpropagation through time; computational cost therefore remains substantial for long horizons.

Parallel studies have demonstrated that carefully designed PINNs or operator networks can reconstruct full-field solutions from very sparse sensors. For instance, the entire cylinder-wake velocity and pressure fields have been reconstructed from sparse sensor measurements collected at only 1% of the entire domain [16], and attention-based Senseiver models have reproduced complex turbulent flows with similarly small data footprints [17]. Nevertheless, these successes have been achieved mainly in regimes of moderate nonlinearity; extrapolation to strongly nonlinear parameter combinations remains an open challenge.

In addition, recent operator-based models such as the Multi-Head Neural Operator (MHNO) efficiently predict full time-series solutions via a single forward pass using time-aware message-passing mechanisms [18]. Unlike residual-based PINNs, Physics-Aware

Recurrent Convolutional neural network (PARC) [19, 20] adopts a dynamical systems approach by embedding diffusion–advection–reaction operators into a recurrent convolutional neural network, enabling accurate simulation of unsteady multiphase flows without explicit PDE residual minimization. These advances underscore a growing demand for real-time surrogates that combine physical fidelity with computational speed.

Following these advances, our group has developed a series of virtual-sensor surrogates that operate in real time: a PINN-based thermal field reconstructor [5], extensions to displacement and stress in solid mechanics [6] and wafer warpage in thermo-elastic settings [7], and, most recently, a causal PINN-Based Surrogate Model (causal PBSM) that incorporates sparse sensors and spatio-temporal coordinates to achieve long-horizon predictions under causal constraints [8]. Although causal PBSM improves long-term accuracy and noise robustness, it still inherits the limitations of explicit time inputs, heuristic causal weighting, and the computational overhead of AD.

To address this issue, we propose the Fixed-Time-Increment PINN-Based Surrogate Model (FTI-PBSM), which enables accurate, real-time full-field reconstruction from sparse data in nonlinear and extrapolative scenarios. A key innovation is that time is removed from the network input altogether; instead, the temporal derivative is obtained numerically by applying fixed-step finite-difference operators to the network output. Replacing chain-rule-based AD with this explicit numerical differentiation (ND) eliminates the need for causal weight tuning and stage-wise scheduling. This redesign not only reduces the training cost and stabilises convergence by decoupling temporal learning from the loss landscape, but also enhances generalization, enabling accurate predictions far beyond the training range.

We validate FTI-PBSM on four benchmark PDEs that span a wide range of physical characteristics: the linear convection equation, diffusion–reaction dynamics, the Korteweg–de Vries (KdV) equation, and the Allen–Cahn equation, thereby covering different levels of nonlinearity, dispersion, and stiffness. Across interpolation and extrapolation settings, and under sparse-sensor configurations, FTI-PBSM consistently outperforms existing causality-enhanced PINN approaches in accuracy, robustness, and scalability.

The remainder of this paper is organised as follows: Section 2 reviews baseline PINN formulations; Section 3 details the FTI-PBSM architecture and training procedure; Section 4 presents numerical results; and Section 5 concludes with a summary and avenues for future work.

## 2. Existing Frameworks for Solving Time-Dependent PDE with PINNs

This section provides an overview of physics informed neural network frameworks for solving time-dependent PDEs. The discussion focuses on two representative approaches, namely the naive PINN [4] and the causal PINN [9], which are designed to reconstruct transient responses based on ICs, BCs, external forcing, and the coefficients of the governing equations.

### 2.1. Naive PINNs

This subsection outlines the naive PINN [4] approach for solving the following time-dependent PDE.

$$u_t + \mathcal{N}[u] = 0, \ x \in \mathbf{\Omega}, \ t \in [0, T] \tag{1}$$

Here, $u = u(t, x)$ denotes the solution, $\mathcal{N}[\cdot]$ is a nonlinear differential operator, $\mathbf{\Omega}$ represents a subset of $\mathbb{R}^D$. The PDE in Eq. (1) is supplemented with corresponding BC and IC:

$$B[u] = h(t, x), \ x \in \Gamma_{BC}, \ t \in [0, T] \tag{2}$$

Here, $B[\cdot]$ denotes a general boundary operator that may represent Dirichlet, Neumann, or Robin-type conditions. The expression $B[u] = 0$ corresponds to the homogeneous form of the boundary condition. While nonzero boundary values can be incorporated as $B[u] = h(t, x)$, where $h$ is a prescribed boundary function, this study presents explicit formulations only for the homogeneous case. The BCs are imposed on the spatial boundary $\Gamma_{BC}$, while the IC is prescribed on the initial time surface $\Gamma_{IC}$ via an initial function $g(x)$, as formulated below:

$$u(0, x) = g(x), \ x \in \Gamma_{IC} \tag{3}$$

For a given input $(t, x)$, the PDE solution is approximated by $\hat{u}$, trained within the PINN framework:

$$u(t,x) \approx \hat{u}(t,x;\hat{\boldsymbol{\theta}}), \tag{4}$$

where $\hat{\boldsymbol{\theta}}$ represents the set of trainable parameters, including the neural network's weights and biases. The approximate solution $\hat{u}(t,x;\hat{\boldsymbol{\theta}})$ is obtained by solving an optimization problem that enforces consistency with the governing PDE as well as the prescribed BCs and ICs. The optimization process seeks to determine the optimal parameter $\hat{\boldsymbol{\theta}}^*$, which is defined as follows:

$$\hat{\boldsymbol{\theta}}^* = \arg\min_{\hat{\boldsymbol{\theta}}} \mathcal{L}_{Total}(\mathbf{X};\hat{\boldsymbol{\theta}}) \tag{5}$$

Here, $\mathcal{L}_{Total}$ is the total loss function and is defined as follows:

$$\mathcal{L}_{Total}(\mathbf{X};\hat{\boldsymbol{\theta}}) = \mathcal{L}_{PDE}(\mathbf{X};\hat{\boldsymbol{\theta}}) + \mathcal{L}_{BC}(\mathbf{X}_{BC};\hat{\boldsymbol{\theta}}) + \mathcal{L}_{IC}(\mathbf{X}_{IC};\hat{\boldsymbol{\theta}}), \tag{6}$$

where $\mathcal{L}_{PDE}$, $\mathcal{L}_{BC}$, and $\mathcal{L}_{IC}$ represent the loss components associated with the PDE, BCs, and ICs, respectively, and are given by:

$$\mathcal{L}_{PDE}(\mathbf{X}_{PDE};\hat{\boldsymbol{\theta}}) = \frac{1}{N_{PDE}} \sum_{i=1}^{N_{PDE}} \left\| \hat{u}_t(\mathbf{X}_{PDE}^i;\hat{\boldsymbol{\theta}}) + \mathcal{N}[\hat{u}](\mathbf{X}_{PDE}^i;\hat{\boldsymbol{\theta}}) \right\|^2, \tag{7}$$

$$\mathcal{L}_{BC}(\mathbf{X}_{BC};\hat{\boldsymbol{\theta}}) = \frac{1}{N_{BC}} \sum_{i=1}^{N_{BC}} \left\| B[\hat{u}](\mathbf{X}_{BC}^i;\hat{\boldsymbol{\theta}}) \right\|^2, \tag{8}$$

$$\mathcal{L}_{IC}(\mathbf{X}_{IC};\hat{\boldsymbol{\theta}}) = \frac{1}{N_{IC}} \sum_{i=1}^{N_{IC}} \left\| \hat{u}(\mathbf{X}_{IC}^i;\hat{\boldsymbol{\theta}}) - g(\mathbf{X}_{IC}^i) \right\|^2. \tag{9}$$

In the domain, $\mathbf{X}_{PDE}$ denotes the set of collocation points used to evaluate the PDE residual, corresponding to coordinates $(t, x)$. $\mathbf{X}_{BC}$ and $\mathbf{X}_{IC}$ denote the collocation points used for enforcing the BCs and ICs, respectively; Similarly, $N_{PDE}$, $N_{BC}$, and $N_{IC}$ refer to the number of collocation points associated with each component of the loss. Gradients of the approximate solution $\hat{u}$ for $(t, x)$, such as $\hat{u}_t$ and $\hat{u}_x$, are computed via Automatic Differentiation (AD) [21]. For higher-order constraints—e.g., second-order derivatives—nested applications of AD are used to compute terms like $\hat{u}_{tt}$ and $\hat{u}_{xx}$, which are directly incorporated into the loss terms $\mathcal{L}_{BC}$ and $\mathcal{L}_{IC}$. The model parameters $\hat{\boldsymbol{\theta}}$ are optimized via gradient descent to minimize the total loss $\mathcal{L}_{Total}$, as defined in Eq. (6). In cases where sensor data are available, a data loss term $\mathcal{L}_{Total}$ can also be included, defined as follows:

$$\mathcal{L}_{data}\left(\mathbf{X}_{data};\hat{\boldsymbol{\theta}}\right) = \frac{1}{N_T \times m} \sum_{i=1}^{N_T} \sum_{j=1}^{m} \left\| \hat{u}\left(t^i, x^j; \hat{\boldsymbol{\theta}}\right) - u_{s_n}\left(t^i, x^j\right) \right\|^2 . \tag{10}$$

Here, $\mathbf{X}_{data}$ denotes the set of sensor data points, where $u_{s_n}(t^i, x^j)$ represents the measurement from the $j$-th sensor at the $i$-th time step. The total number of sensors is denoted by $m$, and $N_T$ indicates the number of time steps.

However, when naive PINNs are trained using a total loss over time-dependent PDEs, they often exhibit convergence difficulties. As the temporal domain extends, residuals near the initial time tend to increase. Consequently, gradient signals associated with enforcing the IC diminish, leading the network to deviate from the correct solution trajectory, converge slowly, or collapse into trivial or non-physical solutions. These issues are further exacerbated in problems involving long-time integration [22] or multiple instances with varying BCs and ICs [23]. The root cause of such instability lies in the absence of an explicit temporal causality in the loss formulation, which hinders consistent enforcement of the IC and degrades solution stability over time. To address these issues, causal training strategies have been introduced to impose temporal causality during loss function minimization.

## 2.2. Causal PINN

To address these issues, causal PINNs [9] have emerged as a promising architectural alternative that explicitly integrates temporal causality into the modeling of dynamical systems. Rather than applying uniform residual minimization across the entire time domain, causal PINNs impose the solution constraints in a progressive manner—starting from the initial time and advancing forward—thus embedding causal structure directly into the training process. This sequential learning strategy enhances convergence and mitigates the risk of trivial solutions, particularly in scenarios involving long temporal horizons or multiple instances. Accordingly, the loss function in Eq. (7) is reformulated as:

$$\mathcal{L}_{PDE}\left(\mathbf{X}_{PDE};\hat{\boldsymbol{\theta}}\right) = \frac{1}{N_T}\sum_{i=1}^{N_T} \lambda^i \mathcal{L}_{PDE}^i\left(t^i;\hat{\boldsymbol{\theta}}\right), \tag{11}$$

where $\lambda^i$ is the weight assigned to the PDE loss at each $t^i$, which is defined as follows:

$$\lambda^1 = 1, \lambda^i = \exp\left(-\epsilon \sum_{q=1}^{i-1} \mathcal{L}_{PDE}^q\left(t^q;\hat{\boldsymbol{\theta}}\right)\right), \text{ for } i = 2, 3, ..., N_T. \tag{12}$$

Here, $\epsilon$ denotes the causality parameter that controls the steepness of the temporal weighting function. A larger $\epsilon$ results in a sharper decay, placing greater emphasis on the initial time steps, whereas a smaller $\epsilon$ yields a more uniform weighting across the time domain. However, selecting an appropriate value for $\epsilon$ is heuristic and problem-dependent, often requiring extensive tuning to balance stability and accuracy.

The term $\mathcal{L}_{PDE}^q\left(t^q;\hat{\boldsymbol{\theta}}\right)$ in Eq. (11) quantifies the PDE residual at the $q$-th time window $t^q$, given by:

$$\mathcal{L}_{PDE}^q\left(t^q;\hat{\boldsymbol{\theta}}\right) = \frac{1}{N_X}\sum_{j=1}^{N_X}\left\|\hat{u}_t\left(t^q, x^j;\hat{\boldsymbol{\theta}}\right) + \mathcal{N}[\hat{u}]\left(t^q, x^j;\hat{\boldsymbol{\theta}}\right)\right\|^2 \tag{13}$$

Here, $N_X$ denotes the number of collocation points within the $q$-th time window $t^q$. A schematic of the causal PINN architecture is provided in Fig. 1.

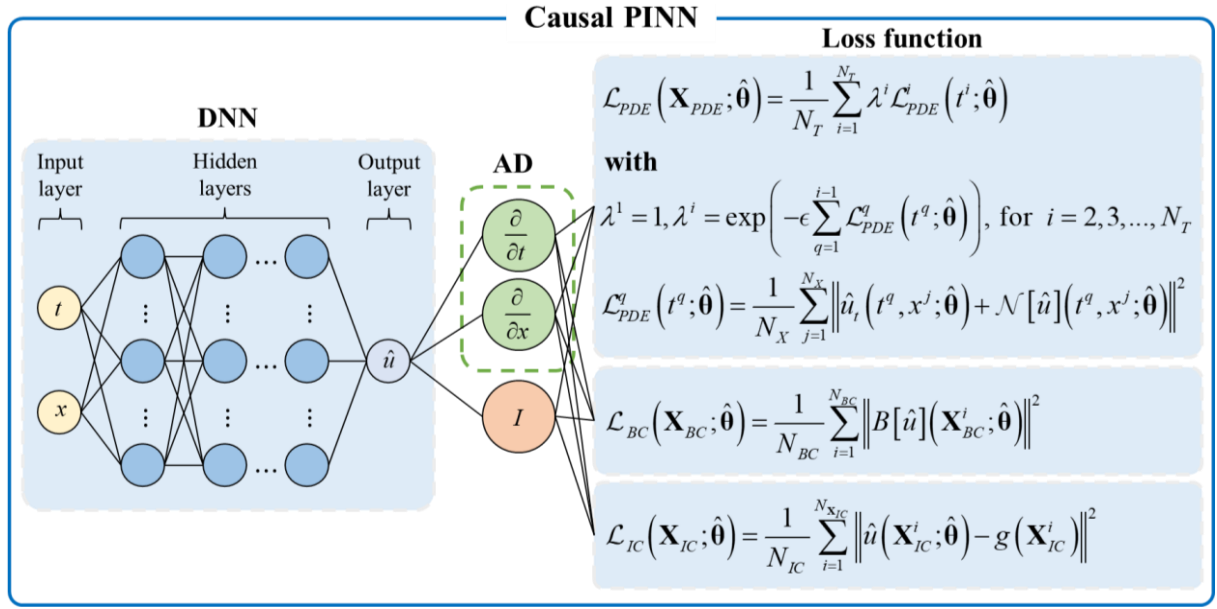

**Fig. 1.** Schematic representation of the causal PINN architecture.

# 3. Proposed Framework: FTI-PBSM

While the causal PINN framework effectively encodes temporal causality and improves convergence for time-dependent PDEs, its application still involves notable challenges. In our previous work, we addressed this by proposing a model that extends the causal PINN into a practical inference framework capable of predicting transient responses from sparse sensor measurements and spatio-temporal coordinates. This model enabled real-time full-field reconstruction across various scenarios and demonstrated improved extrapolation performance compared to naive formulations.

However, the approach exhibited several key limitations. First, time derivatives were computed via AD applied to time-dependent sensor inputs, which often led to unstable training and trivial solutions that minimized the residual loss without capturing meaningful temporal dynamics, even though a remedy was introduced [14]. Second, the model relied on a manually tuned temporal weighting function to enforce causality, which limited its adaptability to diverse temporal behaviors and required problem-specific adjustments.

## 3.1. FTI-PBSM Model Architecture

To overcome these issues, we propose the FTI-PBSM (Fixed-Time-Increment PINN-Based Surrogate Model). As illustrated in Fig. 2, the FTI-PBSM removes the explicit time coordinate from the input features and instead utilizes only spatial coordinates $x$ and time-dependent sensor measurements $\mathbf{u}_s(t) = \left(u_{s_1}(t), u_{s_2}(t), \ldots, u_{s_m}(t)\right)$. The sensor values implicitly encode the current time step, allowing the model to learn dynamic behavior without direct access to time.

Furthermore, the temporal derivatives needed for the physics constraints are computed using Numerical Differentiation (ND) based on a Fixed-Time-Increment (FTI), rather than through AD. This approach improves stability, reduces computational overhead, and enhances the model's ability to generalize across a broad class of nonlinear time-dependent PDE problems.

For the realization of the finite difference scheme, Fig. 3 shows that the collocation points are equidistant in time. Among them, the yellow dots represent sensor points—mandatory for training—as they provide supervision signals and must be included in the collocation set. While temporal points are uniformly spaced to facilitate ND, spatial collocation points may be uniformly or randomly distributed, depending on the problem configuration. Specifically, (a) illustrates a one-dimensional (1D) spatial domain, whereas (b) represents a two-dimensional (2D) spatial domain.

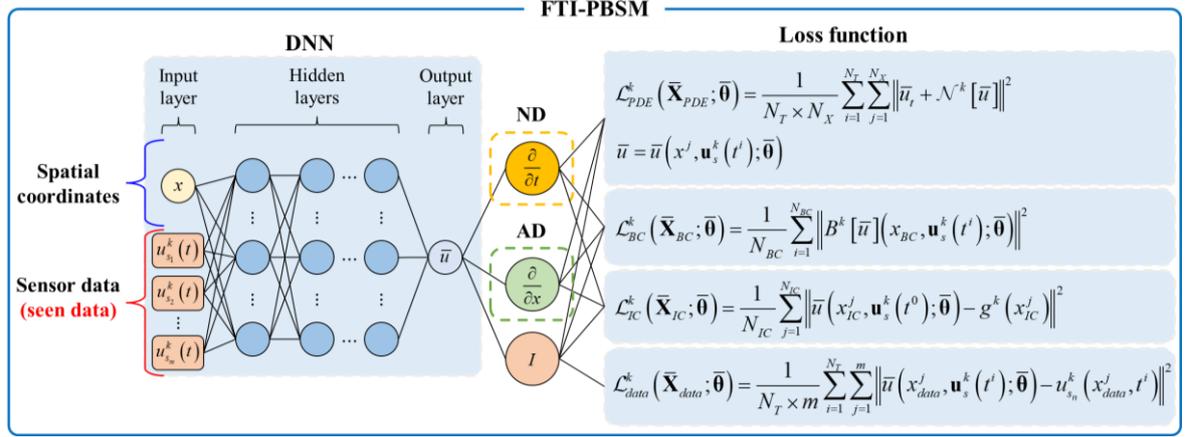

(a) Training phase

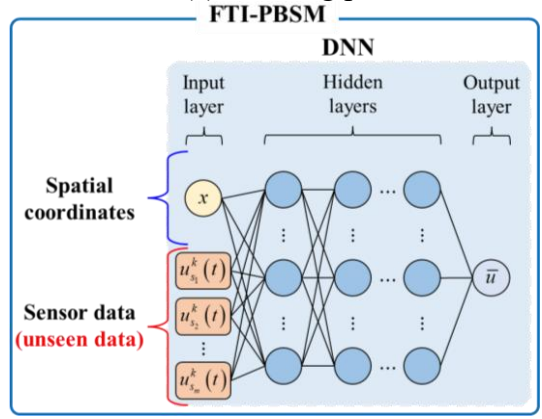

(b) Inference phase

**Fig. 2.** Schematic representation of the FTI-PBSM architecture: (a) Training phase, (b) Inference phase.

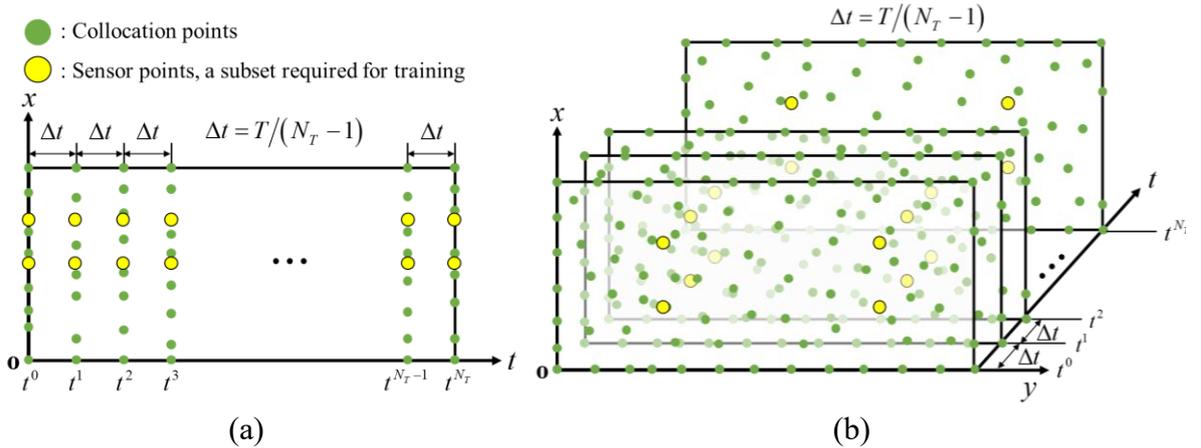

(a)    (b)

**Fig. 3.** Illustration of the collocation point configuration in time. (a) For 1D spatial domains; (b) For 2D spatial domains. Green dots denote temporally equidistant collocation points used for finite difference schemes. Yellow dots indicate sensor points, a subset required for training,

which provide temporal measurements to supervise the model. While time points are uniformly spaced, spatial collocation points may be uniform or random depending on the problem setup.

### 3.2. Temporal Derivative Approximation via Numerical Differentiation

Building on the architecture introduced in the previous section, the FTI-PBSM enforces temporal causality by applying ND directly to the predicted field variable $\bar{u}$, using a finite difference stencil that references adjacent time steps. This strategy not only removes the need for applying the chain rule and AD, but also mitigates the instability that often arises when differentiating explicit time inputs.

The total loss, PDE residual, BC, IC, and data loss for each training sample is formulated as follows:

$$\mathcal{L}_{Total}^{k}\left(\bar{\mathbf{X}};\bar{\boldsymbol{\theta}}\right) = \mathcal{L}_{PDE}^{k}\left(\bar{\mathbf{X}}_{PDE};\bar{\boldsymbol{\theta}}\right) + \mathcal{L}_{BC}^{k}\left(\bar{\mathbf{X}}_{BC};\bar{\boldsymbol{\theta}}\right) + \mathcal{L}_{IC}^{k}\left(\bar{\mathbf{X}}_{IC};\bar{\boldsymbol{\theta}}\right) + \mathcal{L}_{data}^{k}\left(\bar{\mathbf{X}}_{data};\bar{\boldsymbol{\theta}}\right), \tag{14}$$

$$\mathcal{L}_{PDE}^{k}\left(\bar{\mathbf{X}}_{PDE};\bar{\boldsymbol{\theta}}\right) = \frac{1}{N_T \times N_X}\sum_{i=1}^{N_T}\sum_{j=1}^{N_X}\left\|\bar{u}_t\left(x^j,\mathbf{u}_s^k(t^i);\bar{\boldsymbol{\theta}}\right) + \mathcal{N}^k[\bar{u}]\left(x^j,\mathbf{u}_s^k(t^i);\bar{\boldsymbol{\theta}}\right)\right\|^2 \tag{15}$$

with $\bar{\mathbf{X}} = \left(x, \mathbf{u}_s^k(t)\right)$,

$$\mathcal{L}_{BC}^{k}\left(\bar{\mathbf{X}}_{BC};\bar{\boldsymbol{\theta}}\right) = \frac{1}{N_{BC}}\sum_{i=1}^{N_{BC}}\left\|B^k[\bar{u}]\left(x_{BC},\mathbf{u}_s^k(t^i);\bar{\boldsymbol{\theta}}\right)\right\|^2, \tag{16}$$

$$\mathcal{L}_{IC}^{k}\left(\bar{\mathbf{X}}_{IC};\bar{\boldsymbol{\theta}}\right) = \frac{1}{N_{IC}}\sum_{j=1}^{N_{IC}}\left\|\bar{u}\left(x_{IC}^j,\mathbf{u}_s^k(t^0);\bar{\boldsymbol{\theta}}\right) - g^k\left(x_{IC}^j\right)\right\|^2, \tag{17}$$

$$\mathcal{L}_{data}^{k}\left(\bar{\mathbf{X}}_{data};\bar{\boldsymbol{\theta}}\right) = \frac{1}{N_T \times m}\sum_{i=1}^{N_T}\sum_{j=1}^{m}\left\|\bar{u}\left(x_{data}^j,\mathbf{u}_s^k(t^i);\bar{\boldsymbol{\theta}}\right) - u_{s_n}^k\left(x_{data}^j,t^i\right)\right\|^2. \tag{18}$$

Therefore, Eq. (5) is reformulated as follows:

$$\bar{\boldsymbol{\theta}}^* = \arg\min_{\bar{\boldsymbol{\theta}}} \mathcal{L}_{Total}(\bar{\mathbf{X}}; \bar{\boldsymbol{\theta}})$$

with $\mathcal{L}_{Total}(\bar{\mathbf{X}}; \bar{\boldsymbol{\theta}}) = \dfrac{1}{N_s} \sum_{k=1}^{N_s} \mathcal{L}_{Total}^k(\bar{\mathbf{X}}; \bar{\boldsymbol{\theta}})$, (19)

where $N_s$ is the total number of training samples. The formulation of the PDE residual term in Eq. (15) requires temporal derivatives of the predicted field variable at each time step.

Since the FTI-PBSM computes these derivatives through ND, an appropriate time differencing scheme must be selected. Several established methods are available for this purpose, including the Crank–Nicolson (CN) method, Adams–Bashforth (AB) schemes, central difference (CD) methods, and Backward Differentiation Formula (BDF) schemes. Each offers a tradeoff in terms of accuracy, stability, and compatibility with the non-sequential, globally optimized structure of the FTI-PBSM framework.

The CN method is a second-order accurate implicit scheme that evaluates the time derivative by averaging over two adjacent time levels:

$$\left.\dfrac{\partial u}{\partial t}\right|_{t_n} \approx \dfrac{1}{2}\left(f^{n+1} + f^n\right), \qquad (20)$$

Where $f^n = f(u^n, t^n)$ denotes the right hand side of the governing equation in differential form.

This method is well known for its numerical stability and effectiveness in stiff problems. However, it is inherently designed for sequential time marching, where the solution at each time step depends on the solution from the previous step.

In contrast, the FTI-PBSM framework trains the neural network over the entire time domain simultaneously, eliminating step-by-step time marching in the forward pass. Because the model predicts the solution over the whole time domain in a single evaluation, the stability benefits of implicit solvers such as CN do not arise.

Therefore, in this work, we adopt explicit time differencing strategies that are fully compatible with the FTI formulation: the AB method, the CD scheme, and BDF.

The AB method approximates the time derivative using a linear combination of previously evaluated values. Depending on the time step index, different orders of the AB scheme are employed. Specifically, the Euler method is used for the first step, followed by AB-2, AB-3, and AB-4 in subsequent steps. The general formulation for the AB-4 is:

$$\left.\frac{\partial u}{\partial t}\right|_{t_n} \approx \sum_{j=0}^{3} \beta_j f^{n-j}, \quad \text{with } \beta_0 = \frac{55}{24}, \beta_1 = -\frac{59}{24}, \beta_2 = \frac{37}{24}, \beta_3 = -\frac{9}{24}. \tag{21}$$

This sequential switching strategy ensures consistent accuracy across the entire time domain, including near the initial steps where higher-order schemes are not applicable due to lack of sufficient history.

For time steps near the boundaries, where the stencil is not fully defined, AB-3, AB-2, and AB-1 (Euler method) are used accordingly. This staged approach preserves temporal accuracy while accommodating the limited availability of historical data during the early stages of the simulation.

As an alternative, the fourth order central difference (CD-4) method is also considered. This scheme approximates the temporal derivative using a symmetric five point stencil centered at the target time step:

$$\left.\frac{\partial \bar{u}}{\partial t}\right|_{t_n} \approx \frac{-\bar{u}^{n+2} + 8\bar{u}^{n+1} - 8\bar{u}^{n-1} + \bar{u}^{n-2}}{12\Delta t}. \tag{22}$$

This method computes the derivative directly from the predicted field values, and is suitable when temporal information from both past and future time steps is available. For time steps near the boundaries, where the stencil is not fully defined, a second-order central difference scheme is applied:

$$\left.\frac{\partial \overline{u}}{\partial t}\right|_{t_n} \approx \frac{u^{n+1} - u^{n-1}}{2\Delta t}. \tag{23}$$

We also consider the BDF, an implicit multistep method well known for its robustness in stiff systems. In particular, the BDF scheme takes the form:

$$\left.\frac{\partial \overline{u}}{\partial t}\right|_{t_n} \approx \frac{1}{\Delta t}\sum_{j=0}^{k}\alpha_j^{(k)} u_{n-j}, \quad \text{for } k = 1, 2, 3, 4, \tag{24}$$

where $\alpha_j^{(k)}$ are predefined BDF coefficients corresponding to the scheme order $k$. In the case of the fourth-order BDF (BDF-4), the time derivative is approximated by a weighted sum of the current and four previous values:

$$\left.\frac{\partial \overline{u}}{\partial t}\right|_{t_n} \approx \frac{1}{\Delta t}\left(\frac{25}{12}u^n - 4u^{n-1} + 3u^{n-2} - \frac{4}{3}u^{n-3} + \frac{1}{4}u^{n-4}\right). \tag{25}$$

Lower-order schemes (BDF-1 to BDF-3) are used near the initial time steps where full history is unavailable.

The choice between AB, CD, and BDF schemes is treated as a hyperparameter in our framework. All three methods offer high-order accuracy and are fully compatible with the FTI-PBSM structure [24, 25]. The truncation error associated with these schemes is of order $O(\Delta t^4)$, indicating that accurate derivative approximations can be achieved when the $\Delta t$ is properly chosen. Nevertheless, excessively large values of $\Delta t$ may degrade the numerical accuracy and affect the convergence and stability of training.

Although higher-order time differentiation schemes can be employed to further reduce the truncation error, they typically lead to increased computational cost and implementation complexity. This highlights an inherent trade-off between numerical accuracy and computational efficiency that must be considered in the design of the FTI-PBSM.

As illustrated in Fig. 2(b), the trained FTI-PBSM processes spatial coordinates and sensor data from unseen scenarios, enabling real-time inference of full-field responses across the domain.

The overall procedure for training the FTI-PBSM surrogate model is summarized in Algorithm 1. The training is performed in two stages: the Adam optimizer [26] is first used for initial convergence, and the process switches to the L-BFGS optimizer [27] once the total loss falls below a predefined loss threshold $\varepsilon_L$. The hybrid optimization approach was found to be effective for improving the model's convergence stability and final accuracy. Training continues until the step tolerance $\varepsilon_s$, falls below $1\times10^{-10}$; here, $\varepsilon_s$ refers to the maximum change in model parameters between successive iterations, serving as a convergence criterion.

---

**Algorithm 1** Fixed-Time-Increment PINN-Based Surrogate Modeling (FTI-PBSM)

**Input:** Sensor data $\mathbf{u}_s^k(t)$ for $k=1,\ldots,N_s$, Fixed-Time-Increment $\Delta t$, spatial coordinates $x$, Loss threshold $\varepsilon_L$, Step tolerance $\varepsilon_s$, Optimizer settings: Adam, L-BFGS

1: Initialize model parameters $\bar{\boldsymbol{\theta}}$
2: Define total loss: $\mathcal{L}_{Total}(\bar{\mathbf{X}};\bar{\boldsymbol{\theta}}) = \dfrac{1}{N_s}\sum_{k=1}^{N_s}\mathcal{L}_{Total}^k(\bar{\mathbf{X}};\bar{\boldsymbol{\theta}})$
3: **Phase 1** – Adam Optimization
4:     **repeat**
5:         $\bar{\boldsymbol{\theta}} \leftarrow \bar{\boldsymbol{\theta}} - \eta \cdot \nabla\mathcal{L}_{Total}(\bar{\mathbf{X}};\bar{\boldsymbol{\theta}})$
6:     **until** $\mathcal{L}_{Total}(\bar{\mathbf{X}};\bar{\boldsymbol{\theta}}) < \varepsilon_L$
7: **Phase 2** – L-BFGS Optimization
8:     Minimize $\mathcal{L}_{Total}$ until $\varepsilon_s < 1\times10^{-10}$

**Output:** Trained model parameters $\bar{\boldsymbol{\theta}}^*$

---

# 4. Numerical examples

In this section, we evaluate the performance of the proposed FTI-PBSM in predicting full-field transient responses governed by time-dependent PDEs under varying physical parameters. To assess its accuracy, efficiency, and generalization capabilities, we consider four representative benchmark problems: the convection equation, the diffusion–reaction dynamics, the KdV equation, and the Allen–Cahn equation.

All numerical experiments were conducted using MATLAB R2024a [28] on a workstation equipped with an NVIDIA GeForce RTX 4090 GPU (16 GB). Although training time may vary with hardware configuration, inference time remains effectively constant, demonstrating the model's suitability for real-time applications.

A summary of the physical settings and training configurations used for each example is provided in Table 3. Table 4 lists the key hyperparameters for training the FTI-PBSM. To determine the appropriate ND scheme for each benchmark problem, a preliminary training was conducted and the results are summarized in Table 5 and visualized in Fig. 11. Table 6 reports the average inference time across test cases, in comparison with numerical solutions computed using standard MATLAB solvers, thereby demonstrating the practical computational advantage of the proposed surrogate model.

## 4.1. Convection equation

### 4.1.1. Problem definition

The first benchmark problem involves a convection equation with varying physical parameters, designed to assess the performance of the proposed FTI-PBSM framework. Despite being governed by a linear PDE, this example is chosen to evaluate the model's robustness in long-time integration settings, where conventional PINNs often exhibit error accumulation [29]. The governing equation, along with the IC and BCs, is given by:

$$u_t + \gamma u_x = 0, \ (t,x) \in [0,5] \times [0,2\pi],$$
$$u(0,x) = \sin x, \ u(t,0) = u(t,2\pi), \ u_x(t,0) = u_x(t,2\pi), \tag{26}$$

where $\gamma$ is the convection coefficient, assumed to vary across training samples. An analytical solution to the convection equation is available in the form $u(t,x) = \sin(x - \gamma t)$, which serves as a reference for validating the model predictions.

To train the FTI-PBSM, 15 training samples were generated by uniformly varying the $\gamma$ from 1 to 15. Sensor measurements were collected at three fixed locations: $x = \pi/2, \pi, 3\pi/2$, with each sensor. These measurements serve as a latent representation of time within the model. Therefore, the model receives only the spatial coordinate $x$ and time-dependent sensor measurements $\left(x, u_{s_1}(t), u_{s_2}(t), u_{s_3}(t)\right)$ as input, omitting the explicit use of time $t$. For training, a total of 50,601 collocation points (501 in time × 101 in space) were used, discretized uniformly over the domain with an FTI $\Delta t = 0.01$.

The neural network architecture for the FTI-PBSM consisted of five hidden layers with 32 neurons each, using the Sine activation function. Model training followed a full-batch strategy: the Adam optimizer was initially employed until the total loss dropped below $1 \times 10^{-3}$, after which the L-BFGS algorithm was used for fine-tuning and convergence. The CD-4 scheme was selected for ND based on the results summarized in Table 5.

### 4.1.2. Results and Discussion

To evaluate the predictive performance of the trained FTI-PBSM, we tested the model on two representative cases: an interpolation scenario with $\gamma = 4.3$ and an extrapolation scenario with $\gamma = 30$, which lie within and outside the training parameter range, respectively. Fig. 4(a) shows the reconstructed spatio-temporal solution contours under each condition, including clean and noisy (1% Gaussian noise) sensor data. The FTI-PBSM accurately reconstructs the full-field solution in both interpolation and extrapolation tasks, even under noise, demonstrating its robustness and generalization capability.

To further validate temporal prediction accuracy, Fig. 4(b) presents a comparison of the predicted and ground truth solutions at $x = \pi$ over time for the two test cases. The FTI-PBSM closely follows the reference solutions, with minimal degradation in the presence of noise. These results confirm that the model maintains stable performance under both unseen parameters and mild noise levels, making it well-suited for real-time applications in linear transport-dominated problems.

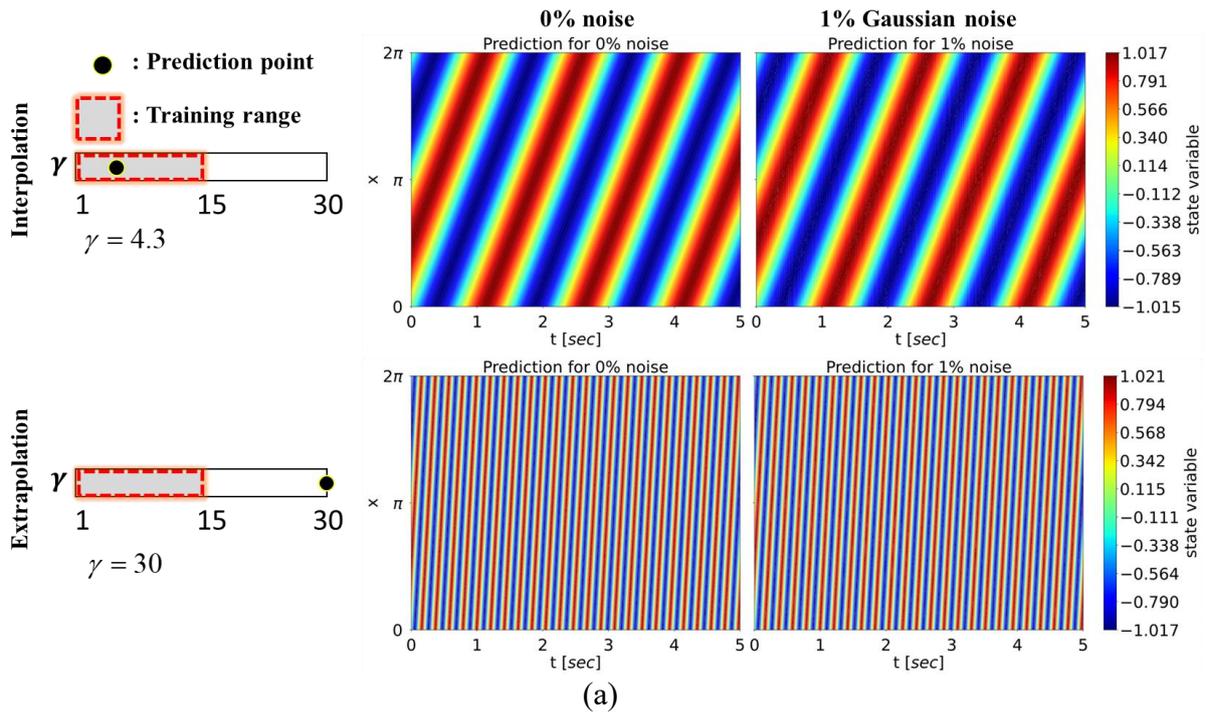

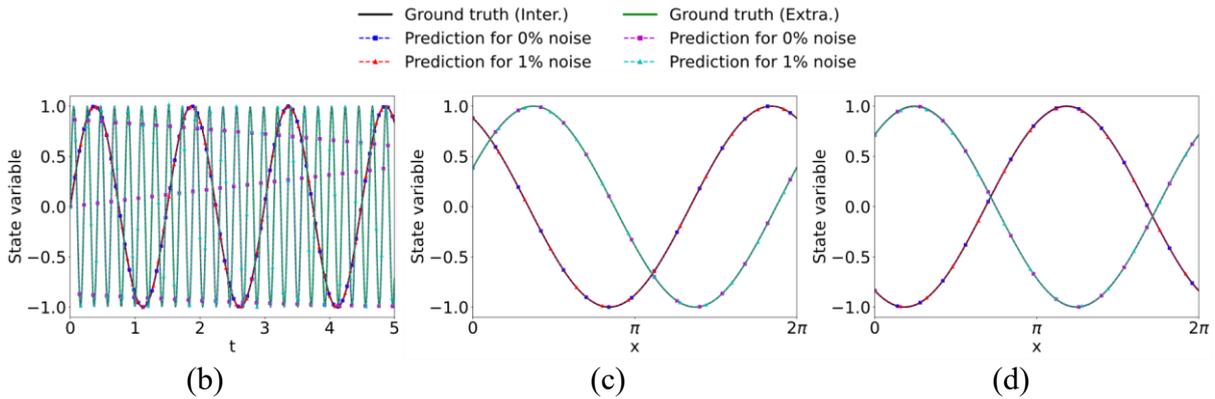

**Fig. 4.** Comparison of predicted spatio-temporal solution fields for the convection equation: (a) Full-field solution contours under interpolation ($\gamma = 4.3$) and extrapolation ($\gamma = 30$) settings, with and without 1% Gaussian noise; (b) Comparison of predicted and ground truth solutions at $x = \pi$; (c) Spatial profile at $t = 2.5$ s; (d) Spatial profile at $t = 5.0$ s.

## 4.2. Diffusion–reaction dynamics

### 4.2.1. Problem definition

The second benchmark problem considers diffusion–reaction dynamics influenced by a time-varying source term $f(x)$ [30]. As depicted in Fig. 5(a), the governing PDE incorporates an unknown diffusion coefficient $D$ and a reaction rate $K$, and is described as follows:

$$u_t = Du_{xx} + Ku^2 + f(x), \ (t,x) \in [0,1] \times [0,1], \tag{27}$$
$$u(t,0) = 0, \ u(t,1) = 0, \ u(0,x) = 0,$$

where $f(x)$ was sampled from a Gaussian Random Field (GRF) [30] and fixed across all training samples, as shown in Fig. 5(b). To evaluate the performance of the proposed FTI-PBSM in capturing complex transient behavior under parametric variation, we generate 25 training samples by varying $D$ and $K$ across five physical cases. In each case, the model is trained to infer the full-field solution $u(t,x)$ given measurements from three fixed sensors located $x = 0.2$, $x = 0.5$, and $x = 0.8$, which provide 101 time-dependent observations, as illustrated in Fig. 5(c).

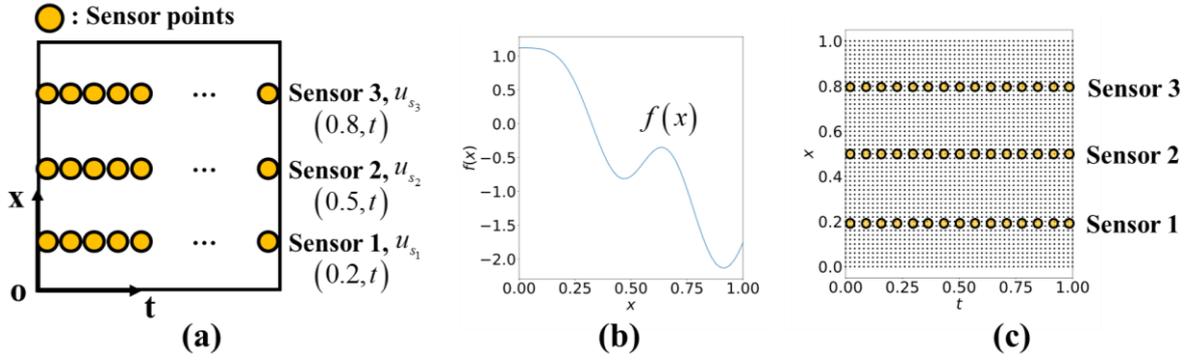

**Fig. 5.** (a) Schematic of the diffusion–reaction dynamics, (b) the time-dependent source term $f(x)$ used in the simulation, and (c) the layout of collocation points and sensor locations for $(t, x)$. Each sensor collects 101 temporal observations of the field variable.

To assess the performance of the proposed FTI-PBSM, this example includes a comparative analysis with the existing causal PBSM framework. While both models aim to reconstruct the full spatio-temporal solution field from sparse sensor measurements, they differ in their treatment of temporal information. Specifically, the causal PBSM takes the spatial and temporal coordinates along with the sensor measurements $(t, x, u_{s_1}(t), u_{s_2}(t), u_{s_3}(t))$ as input and produces the solution $\bar{u}(t, x, u_{s_1}(t), u_{s_2}(t), u_{s_3}(t); \bar{\theta})$ as output. In contrast, the FTI-PBSM receives only the spatial coordinate and time-dependent sensor data $(x, u_{s_1}(t), u_{s_2}(t), u_{s_3}(t))$, omitting the time coordinate $t$ from its input, and generates the corresponding solution field. For training, a total of 5,151 collocation points (101 in time × 51 in space) with a uniform grid were used, with an FTI $\Delta t$ of 0.01.

For the causal PBSM, hyperparameter tuning was conducted for the number of layers, neurons per layer, and the causality parameter $\epsilon$. The final architecture consisted of five layers, each containing 32 neurons, with an $\epsilon$ value of 1.0. The SiLU function [31] was employed as the activation function.

For the FTI-PBSM, the hyperparameters are identical to those of the causal PBSM, except for the omission of the causality parameter, which is not required in this formulation.

Model training for the frameworks utilized a full-batch approach, initially applying Adam optimization until the total loss was reduced to $1\times10^{-3}$, followed by L-BFGS for convergence.

### 4.2.2. Results and Discussion

To assess generalization, both surrogate models were tested on novel parameter combinations of the diffusion coefficient $D$ and reaction rate $K$, covering interpolation and extrapolation scenarios beyond the original 25 training samples. Fig. 6 presents a comparison between the predicted fields generated by the causal PBSM and FTI-PBSM, and the corresponding reference solutions.

As shown in Fig. 6, in interpolation cases, both models successfully reconstructed the overall solution behavior. However, while the causal PBSM exhibited slight deviations from the ground truth in certain regions, the FTI-PBSM achieved near-perfect agreement across the entire domain. This trend became even more evident in extrapolation scenarios: the causal PBSM showed noticeable discrepancies outside the training range, whereas the FTI-PBSM maintained high accuracy and faithfully reproduced the ground truth solution. These results demonstrate that although both models perform reasonably well, the proposed FTI-PBSM exhibits superior precision and generalization capability in both interpolation and extrapolation settings.

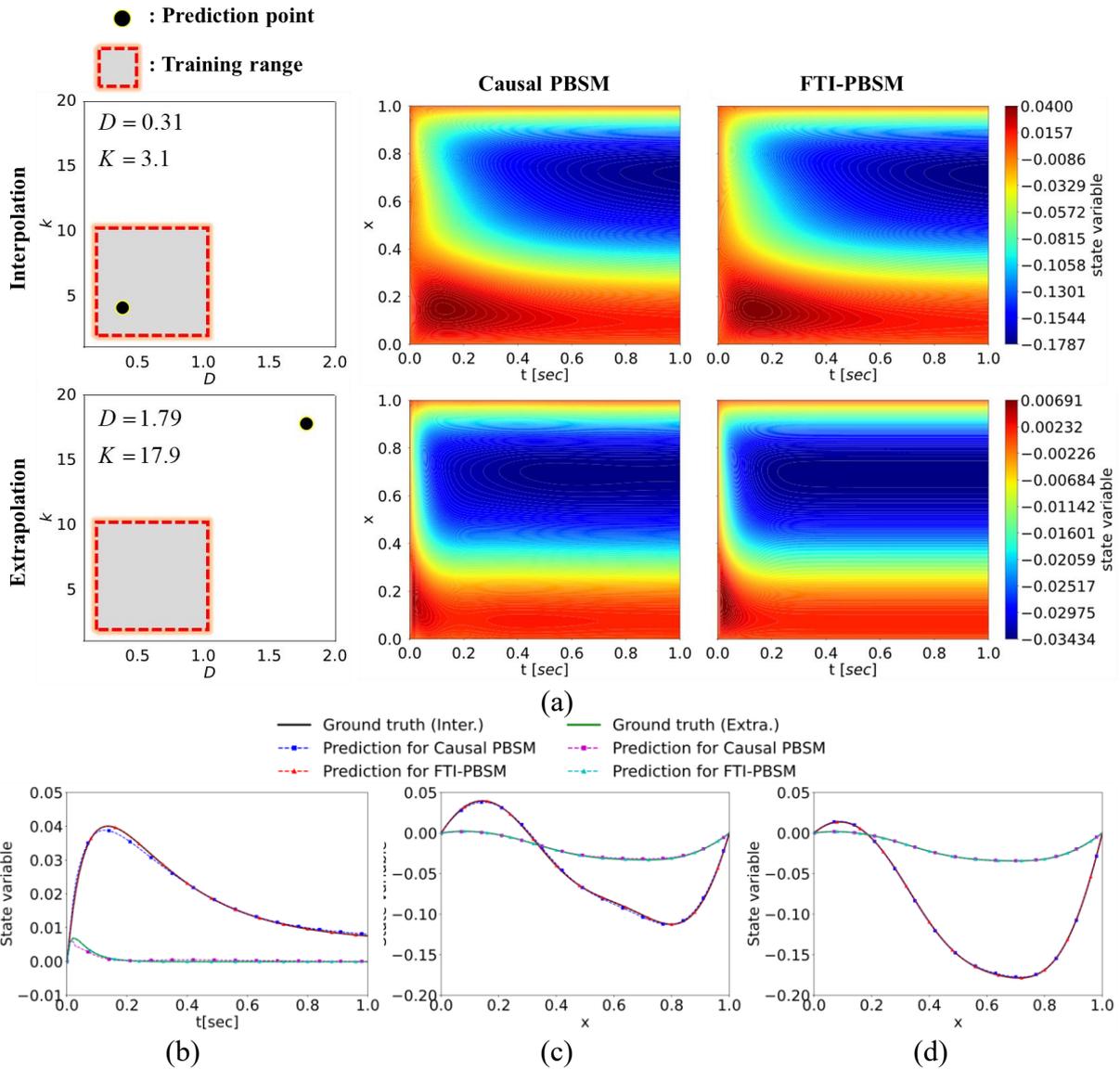

**Fig. 6.** Comparison of predicted spatio-temporal solution fields for the diffusion–reaction dynamics:(a) Predictions of causal PBSM and FTI-PBSM for both interpolation ($D=0.31$, $K=3.1$) and extrapolation ($D=1.79$, $K=17.9$) scenarios; (b) Comparison between the ground truth and surrogate model predictions at $x=0.15$. (b) Comparison of predicted and ground truth solutions at $x=0.16$; (c) Spatial profile at $t=0.16$ s; (d) Spatial profile at $t=1.0$ s.

Additionally, we evaluated and compared the performance of the causal PBSM and the proposed FTI-PBSM in both interpolation and extrapolation tasks, with respect to two key factors: the number of training samples and the presence of measurement noise. To this end, unseen test data were generated by sampling the $D$ and $K$ across extended ranges beyond those used during training. Fig. 7 presents the relative $L_2$ errors for different training sample sizes: 9 (3×3), 25 (5×5), and 100 (10×10), as well as noise levels of 0 percent, 3.0 percent, and 5.0 percent.

When the number of training samples is as low as 9, both models exhibit noticeable errors even within the interpolation region, as highlighted by the red dashed rectangle in Fig. 7, and the accuracy in extrapolation also deteriorates. However, once the sample size increases to 25, both causal PBSM and FTI-PBSM show a clear improvement in predictive accuracy, particularly in extrapolation. Beyond this point, the performance tends to saturate, showing minimal additional gains when increasing to 100 samples. This indicates that the models reach convergence with a moderate number of training cases.

In all cases, the proposed FTI-PBSM consistently outperforms the causal PBSM. It produces more accurate predictions not only within the training domain but also in extrapolated regions, regardless of the number of training samples. As the noise level increases to 3.0 percent and 5.0 percent, both models experience a slight decrease in accuracy, but the overall degradation remains mild. Notably, the FTI-PBSM maintains higher robustness, showing less sensitivity to noise and providing more accurate predictions across all evaluated scenarios.

Table 1 summarizes the training time required for different numbers of training samples. As expected, fewer samples result in faster training, emphasizing the importance of selecting an appropriate number of training cases to balance computational cost and predictive performance. Notably, the FTI-PBSM consistently requires significantly less training time compared to the causal PBSM, due to its simplified architecture that eliminates time-dependent backpropagation. The results suggest that approximately 25 training samples are sufficient to achieve stable and generalizable performance, especially when using the FTI-PBSM.

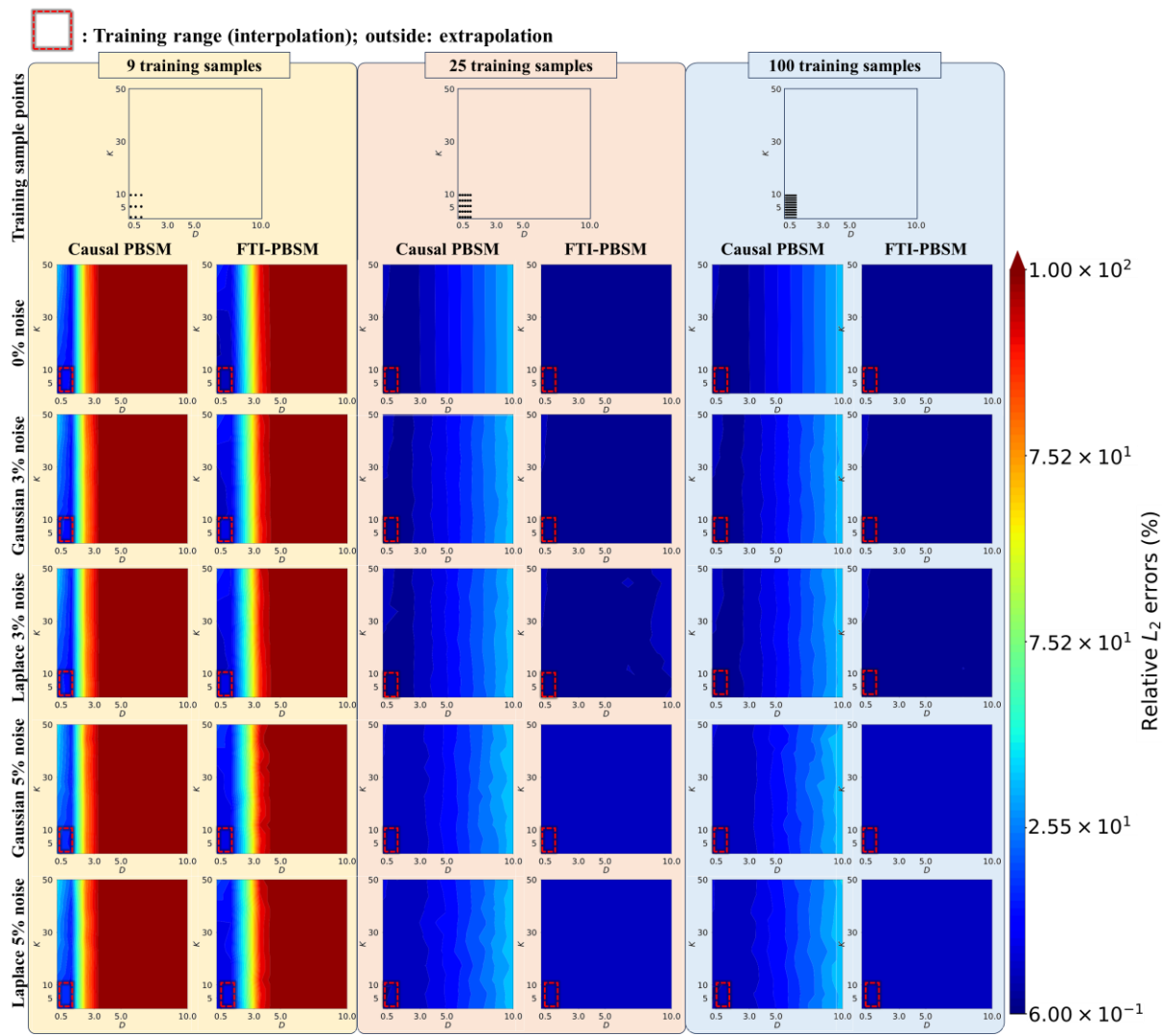

**Fig. 7.** Relative $L_2$ error contours for PBSM predictions on unseen diffusion–reaction data, plotted with respect to varying training sample sizes and sensor noise levels. The red dashed box highlights the region corresponding to the training sample range.

**Table 1.** Training time analysis of different PBSM models under varying sample sizes for the diffusion–reaction dynamics.

| No. Training samples | Training time | |
|---|---|---|
| | Causal PBSM [8] | FTI-PBSM |
| 9 | 13,221 s | **1,484** s |
| 25 | 25,127 s | **2,661** s |
| 100 | 76,119 s | **11,211** s |

To further evaluate the effect of the number of sensors on the performance of the proposed FTI-PBSM, additional experiments were conducted using different sensor configurations: three, two, and one sensor(s). As illustrated in Fig. 5(a), the full configuration uses all three sensors (Sensors 1, 2, and 3), while the two-sensor case uses Sensors 1 and 2, and the single-sensor case uses only Sensor 1.

The model was trained separately under each condition using the same training and test data. Table 2 summarizes the training time and relative $L_2$ errors for 25 unseen data without sensor noise. Although reducing the number of sensors slightly lowers the predictive accuracy, it also results in significantly faster training. This suggests that the number of sensors can be selected based on the desired trade-off between computational cost and accuracy. In particular, using two sensors (Sensors 1 and 2) offers a good balance, maintaining high accuracy while reducing computational overhead.

**Table 2.** Training time and relative $L_2$ error of the FTI-PBSM model evaluated on 25 unseen noiseless test cases, under varying numbers of sensors, for the diffusion–reaction dynamics with unknown diffusion and reaction coefficients.

| No. Sensors used for training | 1 | 2 | 3 |
|---|---|---|---|
| Training time (s) | 1,333 | 2,064 | 2,661 |
| Mean of relative $L_2$ error (%) | 3.121 | 1.171 | 0.510 |

### 4.3. Korteweg–de Vries equation

### 4.3.1. Problem definition

In the third example, we consider the Korteweg–de Vries (KdV) equation with unknown nonlinear and dispersive coefficients. The governing equation, ICs, and BCs are given by

$$u_t + \alpha u u_x + \beta u_{xxx} = 0, \ (t,x) \in [0,5] \times [-1,1],$$

$$u(0,x) = \cos \pi x, \ u(t,-1) = u(t,1), \ u_x(t,-1) = u_x(t,1). \tag{28}$$

Here, $\alpha$ and $\beta$ represent the nonlinear and dispersive coefficients.

To assess the performance of the proposed FTI-PBSM, the model was trained using a dataset generated by varying $\alpha$ and $\beta$ over 25 training samples, as summarized in Table 3.

The training utilized five spatially distributed sensors located at $x = -0.8$, $x = -0.4$, $x = 0.0$, $x = 0.4$, and $x = 0.8$, each collecting 251 temporal measurements. Additionally, 37,901 collocation points (251 in time × 151 in space) were uniformly discretized across the domain to enforce the PDE residual. The FTI-PBSM was configured to receive spatial coordinates and sensor measurements $(x, \mathbf{u}_s(t))$ as inputs, without explicitly including the time coordinate. The network architecture consisted of seven hidden layers with 32 neurons per layer, using the hyperbolic tangent (tanh) function as the activation function. The Adam optimizer was used until the total loss reached $1 \times 10^{-1}$, followed by L-BFGS for convergence refinement.

### 4.3.2. Results and Discussion

The performance of the FTI-PBSM was evaluated on unseen combinations of the nonlinear coefficient $\alpha$ and the dispersive coefficient $\beta$, covering both interpolation and extrapolation settings that were not included in the 25 training samples. As shown in Fig. 8, the model's predictions are compared against the ground truth under various physical conditions.

In the interpolation case ($\alpha = 0.3677$, $\beta = 0.0047$), shown in Fig. 8, the FTI-PBSM accurately reconstructs the full-field solution and captures fine-scale nonlinear wave interactions throughout the domain. In the first extrapolation case (Extra-1, $\alpha = 0.4722$, $\beta = 0.0091$), where the nonlinearity is relatively mild, the model still produces stable predictions and faithfully reproduces the characteristic soliton dynamics of the KdV equation. However, in the second extrapolation case (Extra-2, $\alpha = 0.7800$, $\beta = 0.0038$), where the system becomes highly nonlinear, the model's performance degrades, exhibiting noticeable phase and amplitude discrepancies.

Figs. 8(b)–(d) provide detailed comparisons of the predicted and ground truth responses: (b) temporal evolution at $x = 0$; (c) spatial profile at $t = 2\,\text{s}$; and (d) spatial profile at $t = 5\,\text{s}$. The results demonstrate that while the FTI-PBSM achieves excellent agreement with the ground truth in both interpolation and moderately nonlinear extrapolation settings, its accuracy diminishes in strongly nonlinear regimes.

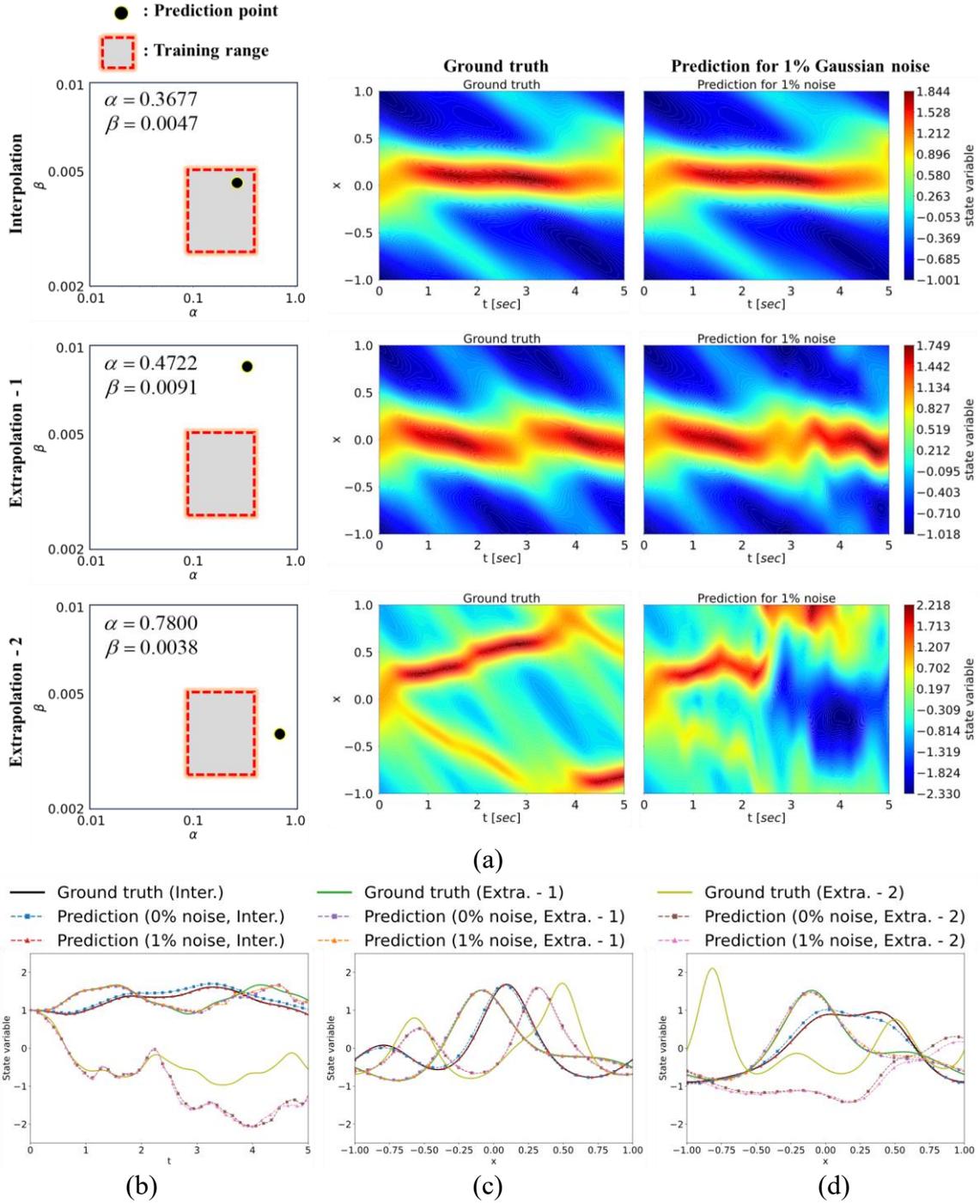

**Fig. 8.** Comparison of predicted spatio-temporal solution fields for the KdV equation: (a) Full-field solution contours under interpolation ($\alpha = 0.3677$, $\beta = 0.0047$) and two extrapolation settings: Extra-1 ($\alpha = 0.4722$, $\beta = 0.0091$) and Extra-2 ($\alpha = 0.7800$, $\beta = 0.0038$), with 1% Gaussian noise; (b) Comparison of predicted and ground truth solutions at $x = 0$; (c) Spatial profile at $t = 2.0$ s; (d) Spatial profile at $t = 5.0$ s.

To more systematically evaluate the model's generalization ability across varying physical regimes, Fig. 9 presents the relative $L_2$ error mapped over the entire $(\alpha, \beta)$ domain. This allows both interpolation and extrapolation performance to be assessed in a unified view. Two training configurations were considered: one using a 3×3 grid (9 samples) and another using a 5×5 grid (25 samples), where training samples were selected by uniformly sampling the nonlinear coefficient $\alpha$ and the dispersive coefficient $\beta$.

When trained with only 9 samples, the error map reveals large deviations primarily in extrapolation regions, while the interpolation region still shows relatively low error. This indicates that even under sparse sampling, the FTI-PBSM can interpolate reasonably well, but its ability to extrapolate to unseen parameter combinations is limited. With 25 training samples, however, the overall error is significantly reduced across the domain, and extrapolation performance improves noticeably—especially in moderately nonlinear regions. In particular, interpolation cases and mild extrapolation settings such as Extra-1 ($\alpha = 0.4722$, $\beta = 0.0091$) exhibit low errors, consistent with the accurate predictions observed in Fig. 9. Despite these improvements, relatively large errors persist in the lower-right corner of the domain, near ($\alpha = 0.7800$, $\beta = 0.0038$). This region is characterized by strong nonlinearity and steep wave interactions, where the model struggles to generalize reliably even with increased training samples.

This trend becomes more pronounced when noise is added to the sensor signals. Although the model remains robust against moderate levels of Gaussian and Laplace noise (3% and 5%), the error increases slightly, reaffirming that the current training distribution lacks sufficient representation of such dynamics.

These observations highlight an important limitation of the current framework: the FTI-PBSM's extrapolation accuracy degrades as the system becomes more nonlinear—particularly in regions with large $\alpha$ and small $\beta$. Future work should aim to improve generalization in these challenging regimes. One promising direction is to selectively augment the training dataset with more samples concentrated in highly nonlinear areas of the parameter space. Additionally, incorporating physics-informed regularization terms that emphasize nonlinear contributions or adopting curriculum learning strategies that gradually expose the model to increasing levels of nonlinearity may further enhance performance. Addressing these issues is expected to extend the applicability of the FTI-PBSM to more complex and strongly nonlinear physical systems.

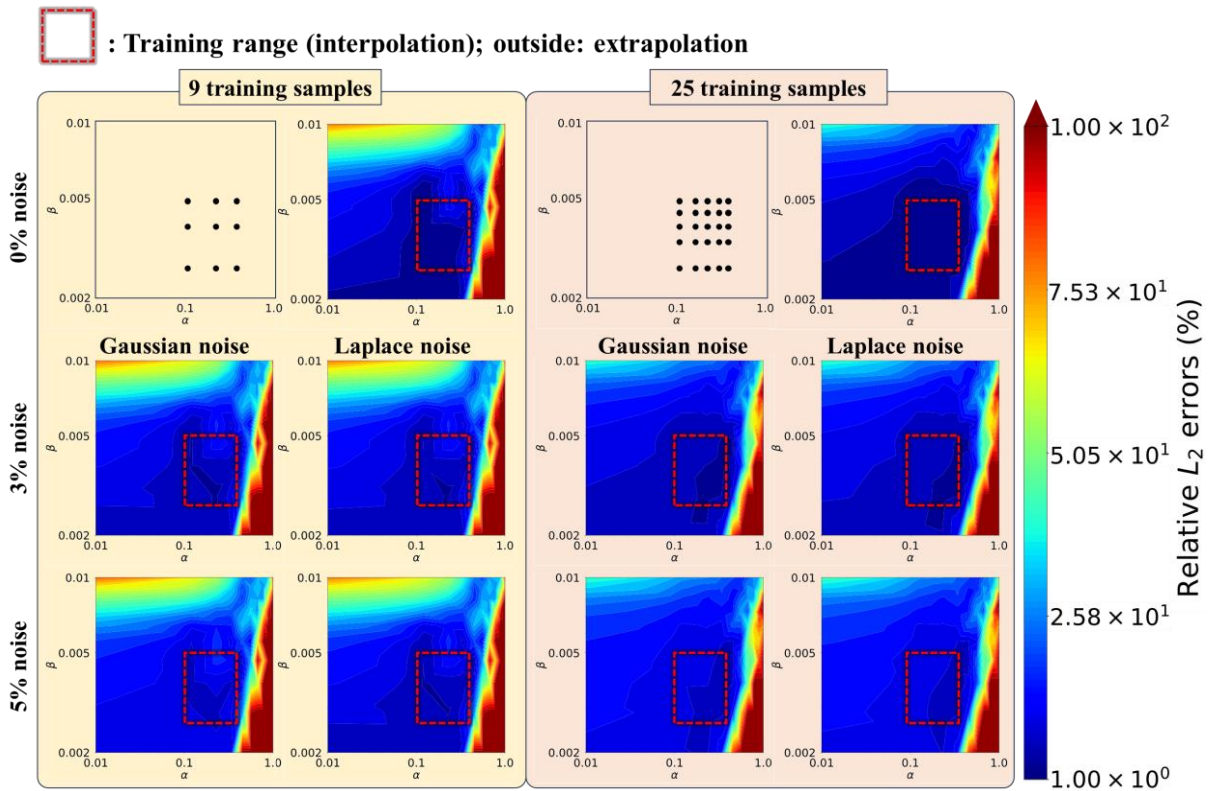

**Fig. 9.** Contour plots of relative $L_2$ errors in FTI-PBSM predictions for the KdV equation with unknown nonlinear and dispersive coefficients, evaluated under varying training sample sizes and sensor noise levels. The red dashed box marks the parameter region covered during training.

### 4.4. Allen-Cahn equation

### 4.4.1. Problem definition

In the last example, we deal with Allen-Cahn equation under unknown diffusion and nonlinear reaction coefficients. The governing equation, BCs, and ICs are as follows:

$$u_t - \delta u_{xx} + \xi u^3 - 5u = 0, \ (t,x) \in [0,2] \times [-1,1],$$

$$u(0,x) = x^2 \cos \pi x, \ u(t,-1) = u(t,1), \ u_x(t,-1) = u_x(t,1), \tag{29}$$

where $\delta$ and $\xi$ are the diffusion and nonlinear reaction coefficients, respectively. These parameters vary across training samples to assess the models' generalization capabilities.

To assess the effectiveness of the proposed FTI-PBSM for modeling the Allen–Cahn system, the model was trained using sensor-based measurements obtained from seven spatial locations: $x = -0.75, -0.5, -0.25, 0.0, 0.25, 0.5,$ and $0.75$. Each sensor recorded 201 temporal samples of the solution variable $u(t,x)$. A total of 25 training samples were generated by varying the $\delta$ and $\xi$, as listed in Table 3. For each training sample, 40,401 collocation points (201 in time × 201 in space) were uniformly discretized across the domain to enforce the PDE residual. The neural network consisted of seven hidden layers with 32 neurons per layer, using the tanh activation function.

### 4.4.2. Results and Discussion

To evaluate the performance of the FTI-PBSM under both interpolation and extrapolation settings, the model was tested on unseen combinations of the $\delta$ and $\xi$ that were not included in the training samples. In addition, to assess the robustness of the model, experiments were conducted with and without Gaussian noise added to the sensor measurements.

Fig. 10 illustrates the comparison results under different conditions. Fig. 10(a) shows the predicted full-field solution contours for interpolation and extrapolation cases, with and without noise. The FTI-PBSM accurately reconstructs the spatio-temporal solution across the entire domain, closely matching the ground truth even in the presence of noise. In interpolation, the predicted fields are nearly indistinguishable from the reference, and in extrapolation, the model maintains strong predictive accuracy, demonstrating its generalization capability.

In the presence of 1% Gaussian noise, some fluctuations are observed near the boundary layer around $x=0$, where the solution exhibits steep spatial transitions. These artifacts are likely caused by the amplification of high-frequency components in the sensor signals during ND. Nevertheless, despite such localized deviations, the FTI-PBSM consistently produces accurate full-field predictions that remain well aligned with the ground truth. The minor impact of the added noise suggests that the model retains stable performance under low-level sensor perturbations.

Figs. 10(b)–(d) present line-wise comparisons of the predicted and reference solutions at selected spatial and temporal slices: (b) shows the temporal evolution at $x=-0.8$, (c) displays the spatial profile at $t=0.5\,\text{s}$, and (d) shows the spatial profile at $t=1.0\,\text{s}$.

In all cases, the FTI-PBSM demonstrates excellent agreement with the ground truth across both noise-free and noisy conditions. While a slight degradation is observed under noise, the overall prediction remains highly accurate and stable. These results confirm that the FTI-PBSM effectively captures the nonlinear reaction–diffusion dynamics and is robust against moderate levels of measurement noise.

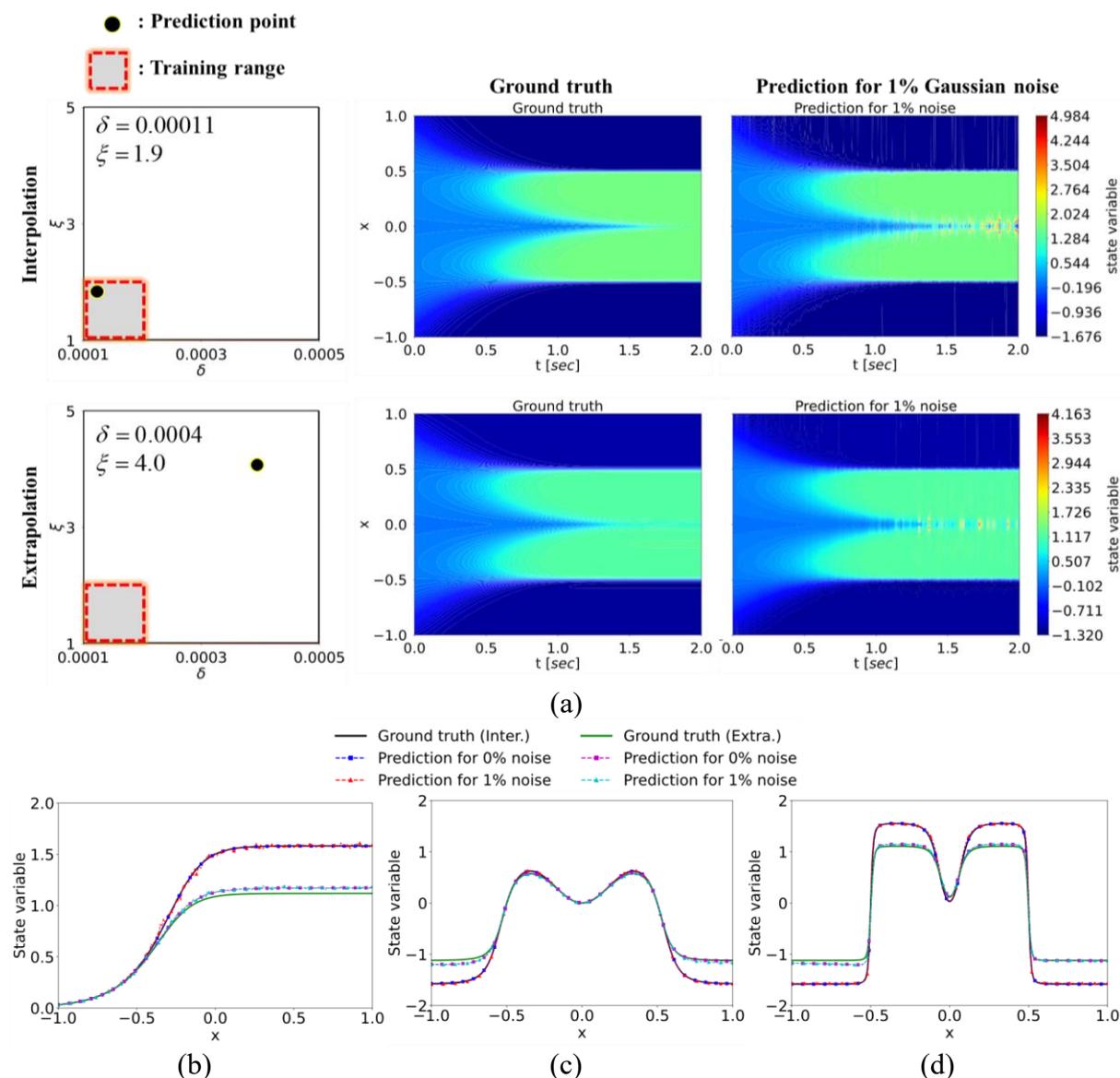

**Fig. 10.** Comparison of predicted spatio-temporal solution fields for the Allen–Cahn equation: (a) Full-field solution contours under interpolation ($\delta = 0.00011$, $\xi = 1.9$) and extrapolation ($\delta = 0.0004$, $\xi = 4.0$) settings, with and without 1% Gaussian noise; (b) Comparison of predicted and ground truth solutions at $x = -0.2$; (c) Spatial profile at $t = 0.5$ s; (d) Spatial profile at $t = 1.0$ s.

## 4.5 Summary

Table 6 summarizes the computational efficiency of each model by comparing their training and inference times against those of conventional numerical solvers, using the training setups from Table 3 and the hyperparameters from Tables 4 and 5. While the proposed FTI-PBSM framework incurs a moderate training cost, it enables ultrafast inference once trained, achieving prediction times within approximately 5 milliseconds per case. This corresponds to up to a 3000 fold speedup compared to traditional solvers in the fastest case, making real-time simulation feasible. Such speed and efficiency highlight the potential of FTI-PBSM for applications in structural anomaly detection as well as broader Prognostics and Health Management (PHM) tasks, even when only a limited number of sensors are available.

To determine the most suitable ND scheme for each benchmark problem, a short preliminary training was conducted using the Adam optimizer for 10,000 iterations. The total loss values were compared among the candidate schemes: AB-4, CD-4, and BDF-4. The scheme that produced the lowest average total loss over the final 100 iterations was selected for the full training process. These selection results are summarized in Table 5, and the corresponding training loss curves for all four benchmark problems are illustrated in Fig. 11. The selected schemes were subsequently adopted in the final hyperparameter settings listed in Table 4.

Nevertheless, it is important to note that the sensor locations used in this study were chosen heuristically. For broader applicability and further performance gains, future work should consider optimizing sensor placement using strategies such as greedy selection or Bayesian optimization.

**Table 3.** Overview of parameter settings and total training sample counts utilized in the four benchmark simulations.

| Example | Parameters | Range | Increment | No. Cases | No. Samples |
|---|---|---|---|---|---|
| Convection equation | $\gamma$ | [1, 15] | $\Delta\gamma = 1.0$ | 15 | 15 |
| Diffusion–Reaction Dynamics | $D$ | [0.1, 1.0] | $\Delta D = 0.18$ | 5 | 25 (=5×5) |
| | $K$ | [1.0, 10] | $\Delta K = 1.8$ | 5 | |
| KdV Equation | $\alpha$ | [0.1, 0.5] | $\Delta\alpha = 0.1$ | 5 | 25 (=5×5) |
| | $\beta$ | [0.003, 0.005] | $\Delta\beta = 0.0005$ | 5 | |
| Allen-Cahn equation | $\delta$ | [0.0001, 0.0002] | $\Delta\delta = 0.000025$ | 5 | 25 (=5×5) |
| | $\xi$ | [1, 2] | $\Delta\xi = 0.25$ | 5 | |

**Table 4.** Overview of the key hyperparameter configurations employed during the training of the FTI-PBSM.

| Example | Spatio-temporal domain $(t,x)$ | Collocation points | FTI $\Delta t$ (s) | LR for Adam optimizer | No. Hidden Layers and Neurons | Activation | Loss threshold $\varepsilon_L$ |
|---|---|---|---|---|---|---|---|
| Convection equation | [0,5]×[0,2π] | 501×101 | 0.01 | 0.001 | 5×32 | Sine | $1\times10^{-3}$ |
| Diffusion–Reaction Dynamics | [0,1]×[0,1] | 101×51 | 0.01 | | 5×32 | SiLU | $1\times10^{-3}$ |
| KdV Equation | [0,5]×[-1,1] | 251×151 | 0.02 | | 7×32 | tanh | $1\times10^{-1}$ |
| Allen–Cahn equation | [0,2]×[-1,1] | 201×201 | 0.01 | | 7×32 | tanh | $1\times10^{-1}$ |

**Table 5.** Numerical differentiation scheme selection results based on preliminary training loss (averaged over the final 100 iterations).

| Example | Candidate Schemes | Average Loss | | | Selected Scheme |
|---|---|---|---|---|---|
| | | AB-4 | CD-4 | BDF-4 | |
| Convection equation | AB-4, CD-4, BDF-4 | $7.27\times10^{-5}$ | **$6.89\times10^{-5}$** | $8.59\times10^{-5}$ | CD-4 |
| Diffusion–Reaction Dynamics | | $3.68\times10^{-5}$ | **$2.43\times10^{-5}$** | $4.20\times10^{-5}$ | CD-4 |
| KdV Equation | | $5.29\times10^{-4}$ | **$4.73\times10^{-4}$** | $5.60\times10^{-4}$ | CD-4 |
| Allen–Cahn equation | | $3.51\times10^{-3}$ | **$2.07\times10^{-3}$** | $3.13\times10^{-3}$ | CD-4 |

**Table 6.** Evaluation of computational performance between MATLAB's built-in numerical solver and the proposed FTI-PBSM across four benchmark cases. (Both computation and prediction times were averaged over 1,000 test samples using a consistent collocation grid of 201 × 101 to ensure fairness in comparison.)

| Example | MATLAB [28, 32, 33] | FTI-PBSM | |
|---|---|---|---|
| | | Training | Prediction |
| Convection equation | - | 8,534 s | 2.58 ms |
| Diffusion–Reaction Dynamics | 0.36 s | 2,661 s | 4.03 ms |
| KdV Equation | 22.6 s | 83,160 s | 6.11 ms |
| Allen–Cahn equation | 2.96 s | 40,441 s | 6.28 ms |

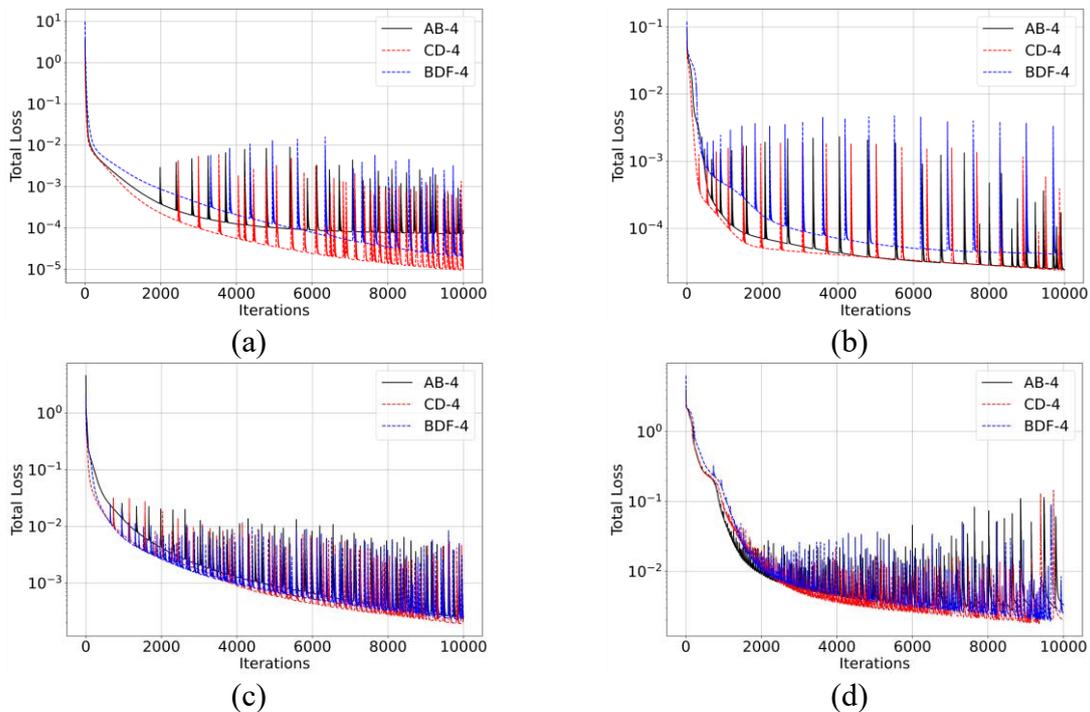

(a) (b) (c) (d)

**Fig. 11.** Comparison of training loss curves for (a) the convection equation, (b) diffusion–reaction dynamics, (c) the Korteweg–de Vries (KdV) equation, and (d) the Allen–Cahn equation using different numerical differentiation (ND) schemes: Adams–Bashforth (AB-4), fourth-order central difference (CD-4), and Backward Differentiation Formula (BDF-4). All methods exhibit stable convergence, with minor variations in convergence rate, under the hyperparameter settings specified in Table 5.

## 5. Conclusion

This study presents the Fixed-Time-Increment PINN-Based Surrogate Model (FTI-PBSM), a new surrogate modeling method for reconstructing transient solutions of time-dependent PDEs in real time, based on sparse measurements. Unlike conventional PINN-based methods that rely on explicit time inputs and Automatic Differentiation (AD), FTI-PBSM captures temporal evolution through higher-order Numerical Differentiation (ND), embedding temporal causality directly into the physics-based loss function. This architectural innovation enables streamlined training, improved stability, and enhanced computational efficiency.

The proposed framework was evaluated on four representative PDE benchmarks—the convection equation, diffusion–reaction dynamics, Korteweg–de Vries (KdV) equation, and Allen–Cahn equation—spanning a wide range of nonlinear, dispersive, and stiff behaviors. The FTI-PBSM exhibited improved predictive accuracy and generalization compared to causal PBSM, maintaining robustness under sparse and noisy sensor conditions across both interpolation and extrapolation regimes. Even with only a small number of sensor measurements, the model accurately reconstructs full spatio-temporal solutions across the entire domain in real time, enabling its deployment in practical environment where dense measurements and computational delays are unacceptable.

In addition to its predictive accuracy, FTI-PBSM features low training computational cost due to the removal of temporal backpropagation and the use of a compact model structure. This computational efficiency suggests that FTI-PBSM can be effectively applied to problems involving faster wave propagation, such as structural vibration and high-Reynolds-number unsteady fluid dynamics.

Future work will aim to extend this framework by incorporating adaptive collocation techniques, optimizing sensor placement strategies, modeling uncertainties in boundary and initial conditions, and validating the method against experimental data for real-world deployment.

**Declaration of Competing Interest**

The authors declare that they have no known competing financial interests or personal relationships that could have appeared to influence the work reported in this paper.

**CRediT authorship contribution statement**

**Hong-Kyun Noh**: Investigation, Methodology, Visualizatoin, Validation, Software, Writing – original draft. **Jeong-Hoon Park**: Investigation, Methodology, Visualizatoin, Validation, Software, Writing – original draft. **Minseok Choi**: Conceptualization, Writing – review & editing. **Jae Hyuk Lim**: Supervision, Conceptualization, Writing – review & editing.

**Acknowledgements**

This work was supported by the Korea Institute of Energy Technology Evaluation and Planning(KETEP) grant funded by the Korea government(MOTIE)(RS-2022-KP002707, Jeonbuk Regional Energy Cluster Training of human resources).